\documentclass[12]{article}
\usepackage{graphics}
\usepackage{epsfig}

\newcommand{\beq}{\begin{equation}}
\newcommand{\eeq}{\end{equation}}

\newcommand{\beqs}{\begin{eqnarray}}
\newcommand{\eeqs}{\end{eqnarray}}

\newcommand{\chose}[2]{\left( \begin{array}{c} #1 \\ #2 \end{array} 
\right)}

\newcommand{\canmat}{\left[ \begin{array}{cc} 0_{6\times 6} & I_{6\times 6} \\ -I_{6\times 6} & 0_{6 \times 6} \end{array} \right]}

\newcommand{\ket}[1]{| #1 \rangle}
\newcommand{\poisn}[2]{\left\{\{ #1 , #2 \}\right\}}
\newcommand{\multi}{e^{\hbar \sum_{\alpha = 1, \beta = 7}^{6 ,12} \omega ^{\alpha \beta }Tr\stackrel{\leftarrow}{\frac{\partial}{\partial A^\alpha }}\stackrel{\rightarrow}{\frac{\partial}{\partial A^\beta }}}}
\newcommand{\st}{\star}
\begin{document}
\bibliographystyle{h-physrev}
\input{epsf}

\title{The Dilatation Operator of $\cal N$ $=$ $4$ SYM and  Classical Limits of Spin Chains and Matrix Models.}
\author{Abhishek Agarwal\thanks{abhishek@pas.rochester.edu} \\
\and
Sarada.G.Rajeev\thanks{rajeev@pas.rochester.edu} \\
University of Rochester. Dept of Physics and Astronomy. \\
Rochester. NY - 14627}
\maketitle

\begin{abstract}
A study of the one loop dilatation operator in the scalar sector of $\cal N$ $=$ $4$ SYM is presented. The dilatation operator is analyzed from the point of view of Hamiltonian matrix models. A Lie  algebra underlying operator mixing in the  planar large $N$ limit is presented, and its role in understanding the integrability of the planar dilatation operator is emphasized. A  classical limit of the dilatation operator is obtained by considering a contraction of this Lie algebra, leading to a new way of constructing classical limits for quantum spin chains. An infinite tower of local conserved charges is constructed in this classical limit purely within the context of the matrix model. The deformation of these charges and their relation to the charges of the spin chain is also elaborated upon.  
\end{abstract}
\section{Introduction:}
In this note we present a systematic study of a Poisson algebraic structure underlying operator mixing in $\cal{N}$$=4$ supersymmetric Yang-Mills theory. In particular we address the question of 'enhancement of symmetry' at the planar limit. We use the formalism developed previously in \cite{Rajeevwilson, rajeev-lee-prl, rajeev-lee-normal} to address this issue. One of the main points that we shall try to make in this paper is that it is impossible to understand the enhanced symmetries  without reference to the underlying Poisson structure. We also construct a classical limit of the dialatation operator, of which the various tractable limits, such as the BMN limit or the planar limit, are deformations. In terms of spin chains, this  amounts to a new way of describing   classical limits of spin chains. The typical problems that one  faces in constructing classical limits of integrable quantum spin chains have to do with finding a set of local conserved charges corresponding to the ones present in the quantum theory, without alterning the form of the Hamiltonian or taking the continuum limit. In the classical limit that we construct in the present paper, such problems do not arise. Moreover, we shall illustrate how a tower of conserved classical charges can be constructed purely within the context of the matrix model without having to make any direct reference to the spin chain or the Bethe-ansatz equations. The passage from this classical limit to the quantum spin chain and vice-versae will be carried out using contractions and deformations of the relevant Poisson algebras.
Since the question is conceptual, rather than computational we shall restrict ourselves to the one-loop dilatation operator in the scalar sector in this paper. 

The one-loop dilatation operator in the scalar sector of $\cal N$ = $4$ SYM, $\Gamma $, takes on the following form \cite{beisert-et-al-conformal}.
\beq
\Gamma = \frac{g^2 }{16\pi ^2}:Tr\left(a^{\dagger i}a^{\dagger i}a_ja_j - a^{\dagger i}a_ja^{\dagger i}a_j +2(a^{\dagger i}a^{\dagger j}a_ja_i - a^{\dagger i}a^{\dagger j}a_ia_j)\right):,\label{dilop}
\eeq
where $a $ and $a ^\dagger $ satisfy the canonical commutation relations,
\beq
\left[ a^{\alpha i}_{j}, a^{\dagger \beta k}{_l} \right] = \hbar \delta^{\alpha \beta}\delta ^{i}_l \delta ^k _j.
\eeq
As a Hamiltonian dynamical system, it is a quantum mechanical matrix model. 
In general, the  dilatation operator of $\cal{N}$$=4$ supersymmetric Yang-Mills theory inherits the  super-conformal invariance of the gauge theory. In the scalar sector, this translates into  an $SO(6)$ invariance, which is the $R$ symmetry of  conformal SYM. At finite $N$ this matrix model is not known to exhibit any integrable features, but in the large $N$ limit, it is equivalent to an $SO(6)$ quantum spin chain, and  a plethora of hidden symmetries  render the quantum spin chain integrable. This remarkable integrability, discovered in the scalar sector by Minahan and Zarembo \cite{minahan-spin}, had its precursor in the study of operator mixing of near BPS operators \cite{bmn, beisert-et-al-bmn, shiraz-et-al, gross-et-al-1, gross-et-al-2}. The dilatation operator in various sectors has since then  been the object of detailed ongoing investigations. The full one loop planar dilatation operator was shown to be an integrable supersymmetric quantum spin chain \cite{beisert-et-al-super-spin}. The higher loop structure of the planar dilatation operator, restricted to a closed sub sector of the scalars,  is also being studied extensively. For  some recent results in this direction see \cite{beisert-et-al-conformal, beisert-et-al-long-range, minahan-et-al-su2}. At higher loops, the conjectured long range spin chain  again turns out to be integrable. This provides an encouraging  indication of a possible all-loops integrability of the theory in the planar limit. This closed   $su(2)$ sub sector of operator mixing, is found  by considering two scalars which are  charged under different $U(1)$ subgroups of the $SO(6)$. Much of what we say in this paper will also be worked out in this sector as well\footnote{It should be kept in mind that the appearance of integrable structures in the study  of the dilatation operator of Yang-Mills theories is not special to the case of $\cal{N}$$=4$. For example, for  studies of  integrable structures associated with operator mixing in Large $N$ QCD, we shall refer the reader to\cite{belitsky-1, belitsky-2, ferretti-zarembo} and references therein.}.

These fascinating discoveries suggest the existance of hidden symmetries within the gauge theory itself. The fact that it is true has been shown by the discovery of Yangian symmetries in superconformal Yang-Mills in \cite{witten-yangian-1, witten-yangian-2}. This discovery provides  further evidence in support of the holographic correspondence\cite{adscft}, as similar symmetries also appear in the corresponding limit of string theory on $AdS^5\times S^5$\cite{hidden-string-symm-1, hidden-string-symm-2}. See also\cite{spin-str-int-st, das-et-al-1, das-et-al-2}. For a detailed comparision of local conserved charges obtained from the integrable structures present in the world sheet sigma model and the dilatation operator of the planar gauge theory, see\cite{gleb-matt-1, gleb-matt-2}. 

However, it is not clear if any of these integrable structures survive beyond the planar large $N$ limit. Non-planar effects in operator mixing have already been investigated in great  detail in the context of the plane-wave gauge theory correspondence. For a nice review on the subject and a more complete list of references, see\cite{bmn-rev}. In the limit of large $R$ charge , it has been shown $\lambda ^{\prime} = g^2_{YM}N/J^2$ plays the role of an effective 't Hooft coupling, and that mixing between single and multi trace operators is unavoidable \cite{shiraz-et-al, gross-et-al-1, gross-et-al-2}, with  $J^4/N^2$ playing the role of an effective genus counting parameter. In this limit, the dilatation operator can be realized as the Hamiltonian of a many body quantum mechanical system\cite{beisert-qm}, which , though tractable in perturbation theory, is not known to be exactly solvable. However, in this approximation, string theory becomes solvable, allowing one to get non-perturbative predictions about the anomalous dimensions of near BPS operators, which are consistent with perturbative gauge theory calculations\cite{bmn, gross-et-al-1, gross-et-al-2} at low orders in perturbation theory. Interesting discrepancies begin to appear at higher orders; see\cite{callan-et-al-1} and references therein for the status of the situation. It is worth mentioning that semi-classical solvability of string theory goes beyond predicting anomalous dimensions for operators which are close to being chiral primaries. Predictions can also be made using the string sigma model for the anomalous dimensions of certain kinds of far from BPS operators \cite{frolov-tseytlin-mul-spin}. For a recent review on this approach see\cite{tseytlin-rev}.  Progress has also been achieved  in the non-planar sector by modelling the splitting of joining of traces within the spin chain approach \cite{bellucci-et-al}.

The picture that has emerged through these investigations suggests that the planar Large $N$ limit enhances the symmetries of the gauge theory in a remarkable way, allowing for the above mentioned integrable structures to simplify the analysis and deepen the understanding of $\cal N$ = $4$ super Yang-Mills. Conversely, the special features of the planar limit must survive as broken or deformed symmetries of the full theory. This is the issue that we address in the present paper.  The main object of our investigation is the matrix model given in (\ref{dilop}), thought of as a Hamiltonian dynamical system.  Related matrix models have also been arrived at by either dimensionally reducing Yang-Mills on $S^3 \times R$\cite{janik-1, bmn, gross-et-al-1, gross-et-al-2} or by looking either at the PP wave limit of matrix theory\cite{keshav-et-al-1, plefka-matrix-1, plefka-matrix-2, plefka-matrix-3}.  For discussions on various symmetry properties of SYM in the plane wave limit and issues related to dimensional reduction, see\cite{das=n=4-1, das=n=4-2,okuyama-1}. String bit models that encode non-planar effects of operator mixing, for example \cite{verlinde-bit-1}, also allude to closely related matrix models. Results obtained by studying the spectrum of such matrix models have yielded in independent tests of results obtained by other techniques, such as perturbation theory or string calculations. However the analysis of these matrix models as Hamiltonian dynamical systems has not received much attention. 

The basic paradigm that we are going to use in this paper is that  large $N$ limits of  matrix models results in a classical theories. To such classical theories are associated relevant Poisson algebras. Such structures were explored  in the context of light cone QCD in \cite{Rajeevwilson} and were realized to be central to the understanding of any Hamiltonian matrix model in \cite{rajeev-lee-prl}. For a review on the this approach towards matrix models see\cite{rajeev-lee-review}. It was shown in this series of papers that the idea that large $N$ limits produce non-trivial classical theories can be made quite precise, and the  limiting classical Hamiltonians and  Poisson brackets can be explicitely constructed. Moreover, a given quantum theory can have many different classical limits, which can be obtained by letting one or more parameters appearing in the quantum theory approach zero while a combination of them is held fixed. In the current problem, there are three natural parameters.\\
{\bf 1:} $g^2 $, which is the  coupling of the gauge theory. This is also the coupling constant of the matrix model. \\
{\bf 2:}The second parameter is the familiar $\frac{1}{N}$. \\
{\bf 3:} The third parameter is $\hbar $ i.e  parameter measuring the quantum fluctuations of the  observables of the dilatation operator, which can legitimately be thought of as the  Hamiltonian of a quantum mechanical system. It is to be distinguished from the Plank's constant appearing in the gauge theory, which has been absorbed in $g^2$. 

Isolating the set of $U(N)$ invariant ovservables that dominate the Large $N$ limit and restricting their action to single trace states generates a Lie algebra obeyed by these observables. We shall refer to it as the planar Lie algebra in this paper. This Lie algebra  can be thought of as a contraction of the full associative algebra of ovservables present in the quantum mechanical problem. In this limit, 
$\lambda = \frac{g^2N}{16\pi ^2}$ and $\hbar $ are held fixed, while $\frac{1}{N}$ approaches zero. This provides a way of thinking about  'quantum' spin chains as  'classical' theories\footnote{This fact can also be established using deformation theoretic techniques, as was done in \cite{Rajeevwilson, rajeev-lee-prl}. We shall briefly review this in the appendix.}\footnote{The fact that conserved charges of the sigma model, computed using classical and semiclassical techniques, match those of the dilatation operator (at one and two loops \cite{gleb-matt-1, gleb-matt-2}) also indicates that 
quantum spin-chins  can have a dual classical description.}. The observables of the classical theory being a subset of the single trace operators of the matrix model and the Poisson bracket being the one implied by the  planar Lie algebra. This Lie algebra is of conceptual importance in the context of integrability, as it provides the correct Poisson algebra {\it with respect to which} the hidden symmetries leading to the integrability of the planar dilatation operator are realized. It also allows for an understanding of the breakdown of the symmetries present at the planar level in terms of quantum  deformations of the relevant Poisson algebra. Apart from addressing this conceptual issue, we shall also work out a further contraction of the Lie algebra in this paper. This contraction shall be achieved by letting $\hbar $ approach zero in the planar Lie algebra. The resulting Poisson structure defines an unconventional, but useful classical limit of quantum spin chains. Its usefulness lies in letting one construct an infinite set of conserved quantities in the classical limit, which are in one to one correspondence with the local conserved charges of the spin chain, without having to take the continuum limit or being forced to modify the Hamiltonian. For a discussion about these issues related to the classical limits of spin chains, we shall refer the reader to \cite{mathieu-charges-1, fadbook}.   Moreover, we shall be able to construct these charges, purely within the context of the matrix model without ever having to refer directly to  the monodromy matrix or the Yang-Baxter equations! Since the spin chain is a quantum deformation of this rather unconventional system, integrability of the spin chain implies that this classical integrability is preserved at the quantum level as well, when quantization is carried out in the sense of $\hbar $.

The paper is organized as follows. In the next section, we shall briefly review  the planar Lie algebra and comment on  the connection of spin chains to planar matrix models. In the section following that,  we shall address the issue of integrability of the quantum spin chains  within the context of the matrix model. After that, we shall work out the contraction of the planar Lie algebra that was referred to above, and construct the classical charges.  We shall end the paper with a short discussion. Some important deformation theoretic results related to contractions of the associative algebra of observables of the quantum theory are gathered together in the appendix, along with some comments about the operator product expansion of the gauge theory.

\section{The Poisson Algebra and The Dilatation operator:}
In this section and the following subsection, we shall briefly review the results necessary for the understanding the main issue explained in the introduction. 
Let us first recapitulate some general notions about operator mixing in the scalar sector. 
The dilatation operator can equivalently be written as \cite{beisert-et-al-conformal},
\beq
\Gamma = \frac{\lambda }{32\pi ^2N}Tr\left(2\left[a^{\dagger i},a^{\dagger j}\right]\left[a_j,a_i\right] - \left[a^{\dagger i},a_j\right]\left[a^{\dagger i},a_j\right]\right).
\eeq
This second form, being a product of commutators, makes it obvious that the dilatation operator vanishes in the case of a $U(1)$ theory. Normal ordering shall be implied throughout the paper, and we shall not specify it explicitely unless required. In the matrix model, $i,j $ range from 1 to 6. Operator mixing for the scalars is now understood in the following way. Local operators in the gauge theory formed out of the scalars are realized as states of this matrix model. A general multi-trace operator in the scalar sector,
\beq
\Upsilon ^{I,J,\cdots, M} = Tr(\Phi ^{i_1} \cdots \Phi ^{i_{|I|}})Tr(\Phi ^{j_1} \cdots \Phi ^{j_{|J|}})\cdots Tr(\Phi ^{m_1} \cdots \Phi ^{m_{|M|}}),
\eeq 
corresponds to the state,
\beq
\Upsilon ^{I,J,\cdots, M} \mapsto \ket{I,J, \cdots, M} = O^I O^J \cdots O^M\ket{0},
\eeq
where,
\beq
O^I = \frac{1}{\sqrt{N^{|I| -2}}}Tr \left(a^{\dagger i_1 }\cdots a^{\dagger i_n }\right).
\eeq
We have denoted ordered strings of indices by capital letters, for example, \\$\{i_1, i_2, \cdots ,i_{|I|}\}$ $= I$, while $|I|$ denotes the number of bits present in the string.
Keeping in mind that the dilatation operator is closely related to the string field Hamiltonian, this mapping of operators to states seems is a 
reflection of the operator-state correspondence between local gauge invariant operators of the gauge theory and states of the string theory. This is most explicit in the plane wave limit, where $\Gamma -J$ is indeed the string field Hamiltonian, $J$ being the $R$ charge generator.

The matrix model $\Gamma $ is dynamical, i.e, we are dealing with the quantum mechanics of six Hermitian matrices. The 'time' in this quantum mechanical problem has a natural interpretation as the 
logarithm of the scale of the gauge theory.
In fact, the Schrodinger equation for the time evolution of the states 
\beq
i\hbar \frac{\partial }{\partial t}\ket{I_i \cdots I_n} = \Gamma \ket{I_i \cdots I_n},
\eeq
has the same information as 
the Callan-Symanzyk equation for the corresponding gauge theory operators.
\subsection{The Large $N$ limit and the Planar Lie Algebra:}
To leading order in $\frac{1}{N}$, the only relevant operators are the ones that do not split such single trace states in two. These operators are of the following type,
\beq
O^I_J = \frac{1}{\sqrt{N^{|I| +|J| -2}}}Tr \left(a^{\dagger i_1 }\cdots a^{\dagger i_n }a_{j_m }\cdots a_{j_1 }\right),
\eeq
for which normal ordering in the (sense of operator ordering) is compatible with the ordering implied by matrix multiplication. Note the reversal of order on the lower indices of the operator. We shall see that this is a useful convention to use.

The action of these operators on single trace states, to leading order in $\frac{1}{N}$ is,
\beq
O^I_J\ket{K} = \lim_{N \rightarrow \infty } \left(O^I_J \st \frac{1}{\sqrt{N^{|I|}}}Tr a^{\dagger k_1 }\cdots a^{\dagger k_n }\right) \ket{0} = \sum_{K_1K_2 =K}\delta ^{K_1}_J \ket{IK_2},
\eeq
where the  sum is over all ways of splitting the string of indices $K$ into two subsets $K = K_1K_2$ such that the first subset is equal to $J$. $K_2$ can be the empty set. The star product, referred to in this equation has been explained in the appendix. Looking at the antisymmetric part of the action of two such  operators generates the planar Lie algebra, which was derived in \cite{rajeev-lee-prl}. Written out in complete detail it reads as follows.
\beqs
\{O^I_J, O^K_L\}_{2}=(\hbar )^{|K|}(\delta ^K_ JO^I_L + \sum_{J_1J_2 =J}\delta ^K_ {J_2}O^I_{J_1L} + \sum_{J_1J_2 =J}\delta ^K_ {J_1}O^I_{LJ_2}\nonumber \\
+\sum_{J_1J_2 J_3=J}\delta ^K_ {J_2}O^I_{J_1LJ_3} + \sum_{\stackrel{J_1J_2 J_3=J}{K_1K_2=K}}\delta ^{K_1}_ {J_3}\delta ^{K_2}_ {J_1}\tilde{O}^I_{J_2L})\nonumber \\
+(\hbar )^{|J|}(\sum_{K_1K_2 =K}\delta ^{K_1}_{J}O^{IK_2}_{L} + \sum_{K_1K_2 =K}\delta ^{K_2}_{J}O^{K_1 I}_{L} + \sum_{K_1K_2K_3 =K}\delta ^{K_2}_{J}O^{K_1IK_3}_{L}\nonumber\\
+\sum_{\stackrel{J_1J_2=J}{K_1K_2=K}}\delta ^{K_1}_ {J_2}\delta ^{K_2}_ {J_1}\tilde{O}^I_{L} + \sum_{\stackrel{J_1J_2=J}{K_1K_2K_3=K}}\delta ^{K_1}_ {J_2}\delta ^{K_3}_ {J_1}\tilde{O}^{IK_2}_{L})\nonumber \\
+\sum_{\stackrel{J_1J_2=J}{K_1K_2=K}}(\hbar )^{|K_1|}\delta ^{K_1}_ {J_2}O^{IK_2}_{J_1L} + \sum_{\stackrel{J_1J_2=J}{K_1K_2=K}}(\hbar )^{|K_2|}\delta ^{K_2}_ {J_1}O^{K_1I}_{LJ_2}\nonumber \\
+\sum_{\stackrel{J_1J_2J_3=J}{K_1K_2K_3=K}}\hbar ^{J_1 + J_3}\delta ^{K_1}_ {J_3}\delta ^{K_3}_ {J_1}\tilde{O}^{IK_2}_{J_2L} - \chose{I \Leftrightarrow K}{J \Leftrightarrow K}.\nonumber \label{planar-lie}
\eeqs

In the above equation, 

\beq
\tilde{O}^{I}_{J} = O^I_J - \sum_{i=1}^{n} O^{kI}_{kJ}.
\eeq
This Lie algebra is best understood in a diagrammatic fashion for which we refer the reader to \cite{rajeev-lee-review}. This Lie bracket can also be thought of as a Poisson bracket between the  observables of the matrix model. Hence, there are two alternative but equivalent ways of looking at the planar limit. It can either be thought of as a quantum spin chain, where, we identify the single trace states of the matrix model with those of the spin chain, and the matrix model observables of the kind mentioned above with corresponding spin chain operators. We shall comment in more detail on this connection in the next section. The other way of looking at the same system is as a classical limit of the matrix model. 

\section{Contractions, Spin-Chains and Integrability:}
Once the contraction to the planar level is carried out single trace states form a complete set.  Moreover, the action of these normal ordered operators on the oscillator states always involves commuting  consecutive annihilation operators (present in the definition of the normal ordered operators) through consecutive creation operators (present in the definition of the state) , i.e. one  has nearest neighbor interactions. This allows for a map between the normal ordered operators and Hamiltonians of quantum spin chains, which is simplest if one interprets  the oscillator states as states of a quantum spin chain.  Let us now translate some important spin chain operators in the language of matrix models.\\
{\bf 1::}The Weyl operator $W^i_j(k)$, which can be regarded as the building block of all local spin chain operators  is such that,
\beq
W^i_j(k)\ket{l_1 \cdots l_{k-1} l_k l_{k+1} \cdots l_m} = \delta ^{l_k}_j \ket{l_1 \cdots l_{k-1} i l_{k+1} \cdots l_m}
\eeq
Similarly, one may have two body operators, operators acting on the spin chain that bring about nearest neighbor interactions. These operators are naturally labeled by the links rather than sites. Of particular interest are three operators. \\
{\bf 2:} The Identity operator $I$.
\beq
I(l)(i_l \otimes i_{l+1}) = (i_l \otimes i_{l+1})
\eeq
{\bf 3:} The Permutation operator $P$.
\beq
P(l)(i_l \otimes i_{l+1}) = (i_{l+1} \otimes i_{l})
\eeq
{\bf 4:} The Trace operator $K$.
\beq
K(l)(i_l \otimes i_{l+1}) = \delta _{i_l  i_{l+1}}\sum _k(i_{k} \otimes i_{k})
\eeq
Thinking of the spin chains as oscillator states, one can see that,
\beq
\sum_{sites} W^i_j(k) \equiv O^i_k,
\sum_{sites} I(l) \equiv O^{ij}_{ij}, \sum_{sites} P(l) \equiv O^{ij}_{ji}, \sum_{sites} K(l) \equiv O^{ii}_{jj}
\eeq
More generally, one can regard most local spin chain operators of interest as being built out of products of Weyl operators  acting on neighboring sites. Hence it is useful to keep in mind the map between a local operator formed out of  products of neighboring Weyl operators and the corresponding matrix model observable, which is\cite{rajeev-lee-prl} 
\beq
\sum_{sites} W^{i_1}_{j_1}(l) W^{i_2}_{j_2}(l+1)\cdots W^{i_b}_{j_b}(l+b-1) \equiv O^{i _1 \cdots i_b}_{j_1 \cdots j_b}.
\eeq
\subsection{Integrability of the Spin Chains from Matrix Models:}

This map between large $N$ matrix models to spin chains is a very generic feature. A natural question to ask is how the question of integrability of the spin chains  posits  itself in the matrix model language. A spin chain Hamiltonian is integrable if it is part of a hierarchy of conserved charges following from the solution of a quantum Yang-Baxter equation. The basic idea is  the following (for a detailed and insightful review on the subject we refer the reader to \cite{faddeev-1}). One starts with a vector space $V$, which is implicitly assumed to be vector space associated with the values of the spin. The starting point is the existence of a $R$ matrix, which is an element of $End V\otimes V$, satisfying the Yang-Baxter relation on $ V\otimes V \otimes V$,
\beq
R_{12}(\lambda )R_{13}(\lambda +\mu )R_{23}(\mu ) = R_{23}(\mu )R_{13}(\lambda +\mu )R_{12}(\lambda )
\eeq
The indices in $R$ reflect the vector spaces in the triple tensor product on which $R$ acts, its action being the trivial one in the remaining vector space. The arguments of $R$, i.e $\lambda, \mu $ refer to the values of the spectral parameter. The simplest non-trivial Hierarchy for this discussion is the one containing the spin half Heisenberg $XXX$ model and we shall use that to illustrate the basic ideas in the remainder of the subsection.  From the point of view of the dilatation operator, this hierarchy is relevant if one restricts oneself to the closed sub sector of two charged scalars \cite{beisert-et-al-conformal, beisert-et-al-long-range, minahan-et-al-su2}.  In this case $V=C^2$, a two dimensional complex vector space. The $R$ matrix for this Hierarchy is,
\beq
R_{ij}(\lambda ) = \lambda I_{ij} + i P_{ij},
\eeq
where $I$ and $P$ are the identity and permutation operators introduced previously.  Typically, $R$ at a shifted value of the spectral parameter is identified with the so called Lax operator.
\beq
R_{ij}(\lambda - \frac{i}{2}) = L_{ij}(\lambda )
\eeq
For a spin chain with $n$ sites, one typically considers a $n+1$ fold tensor product, all but where the last vector space, usually referred to as the auxiliary space, are thought of as the vector spaces making up the Hilbert Fock space $C^2 \otimes \cdots \mbox{n times} \cdots \otimes C^2$ of the spin chain. One then proceeds to construct the monodromy matrix,
\beq
T_{n,a}(\lambda ) = L_{n,a}(\lambda )L_{n-1,a}(\lambda )\cdots L_{1,a}(\lambda ),
\eeq
where $a$ refers to the auxiliary space. the transfer matrix can be written as a matrix acting on the auxiliary space, i.e.
\beq
T_{n,a} = \left(
\begin{array}{cc}
A_n(\lambda )              &  B_n(\lambda )              \\
C_n(\lambda )              &  D_n(\lambda )     \\
\end{array}
\right)
\eeq
with the entries being understood as operators acting on the Hilbert space. Tracing this matrix over the auxiliary space produces a generating function for an infinite set of conserved charges  with the Hamiltonian being one of them. 
Or in other words,
\beq
G(\lambda ) = A(\lambda ) + D(\lambda ) : [G(\lambda ), G(\mu )]=0
\eeq
The coefficients of the expansion of the trace of the monodromy matrix in powers of the spectral parameters are the conserved charges of the quantum spin chain. 
\beq
G(\lambda ) = \sum Q_l\lambda ^l
\eeq
For the particular choice of R matrix, the first non-trivial conserved charge is the permutation operator, 
\beq
\frac{i}{2}\frac{d}{d\lambda } \ln G(\lambda ) = \sum _{links} P(l).
\eeq
The Hamiltonian for the $XXX$ model can now be obtained by adding to this charge, the trivial conserved quantity, i.e the identity operator.
\beq
H_{XXX} = \frac{1}{2}\sum_{links}(P(l) - I(l))
\eeq
Before moving on, it is worth noting that the  next conserved charge in this Hierarchy is,
\beq
H_2 = \sum_{links}(P(l)P(l+1) - P(l+1)P(l)).\label{second-charge}
\eeq
From the previous discussion on the translation of spin chain Hamiltonians to matrix models, we can immediately  write down the corresponding matrix model Hamiltonian;
\beq
H_{XXX-Matrix} = \frac{1}{2N}Tr\left(a^{\dagger i}a^{\dagger j}a_ia_j - a^{\dagger i}a^{\dagger j}a_ja_i \right) = -\frac{1}{4N}Tr\left([a^{\dagger i},a^{\dagger j}][a_j,a_i]\right),
\eeq
where the indices on the creation and annihilation operator now take on two values, corresponding to the spins being up or down.

Let us now address two obvious problems that show up in passing from the 'microscopic' description of the spin chain in terms of sums over permutation and identity operators acting on links to the 'macroscopic' description in terms of the matrix model.
 
The way the translation of the spin chain to the matrix model was achieved in the previous discussion, it might seem that all the microscopic information about the spin system is lost, i.e, there is apparently nothing in the matrix model Hamiltonian that might be interpreted as a spin operator 'at a link', e.g $P(l)$ or $I(l)$. It is only the sum over all the links that appears as a matrix model. But all questions related to the integrability of the spin systems are encoded in the Yang-Baxter relation, which is formulated only in terms of the microscopic variables. Hence the question that needs to be asked is, what information in the matrix model Hamiltonian might inform us that its large $N$ limit is an integrable spin chain? This question is a particularly relevant one in the context of the Dilatation operator, which is a bona-fide matrix model.

It is not difficult to answer this question. The 'site dependent' information is encoded in the coupling constants of the matrix model, which  in reality is a tensor. The particular Matrix model Hamiltonian appearing above can be written in the form,
\beq
H_{XXX-Matrix} = - \frac{1}{4N}\Psi^{kl}_{ij} Tr\left([a^{\dagger i},a^{\dagger j}][a_l,a_k]\right), 
\Psi^{kl}_{ij} = \delta ^k_j\delta ^l_i - \delta ^k_i\delta ^l_j.
\eeq
This tensor, which reflects the $SU(2)$ symmetry of the model, can be thought of as a bona-fide operator on $C^2 \otimes C^2$. To motivate this, it is best to pass to this dual picture, and realize the spin-system matrix model map there. Let us consider the leading $\frac{1}{N}$ action of a  matrix model operator on a two cite spin chain in the planar limit.
\beq
\frac{1}{N^2}\Psi^{kl}_{ij} Tr\left(a^{\dagger i}a^{\dagger j}a_la_k\right)\Xi_{i_1i_2}Tr\left(a^{\dagger i_1}a^{\dagger i_2}\right)|0\rangle
\eeq
The action of the matrix model operator on this state can be translated in to an action of the tensor characterizing the operator on the tensor characterizing the state as 
\beq
\left(\Psi \Xi \right)_{ij} = \Psi^{kl}_{ij}\Xi_{kl} 
\eeq
Regarding $\Psi $ as an element of  $C^2 \otimes C^2$, we see that it is entirely reasonable to think of the tensors characterizing the matrix model Hamiltonians as the site dependent objects of the spin system, and the tensors characterizing the single trace states can literally be identified with the spin chains. If follows in an obvious fashion, that, the permutation and the identity operator correspond to
$\Psi^{kl}_{ij}=\delta ^k_j\delta ^l_i$ and $\Psi^{kl}_{ij}=\delta ^k_i\delta ^l_j$ respectively. Looking at the action of the matrix model operator on a chain with more than two sites, transforms $\Psi$ into a spin chain operator acting on the spin chain with nearest neighbor interactions.
\beq
\left( \Psi \Xi \right)_{i_1 \cdots i_n} = \sum _l \Psi^{pq}_{i_l i_{l+1}}\Xi _{i_1 \cdots i_{l-1}pq i_{l+1} \cdots i_n}.
\eeq
We now see that a matrix model will produce an integrable spin chain if the tensor characterizing the Hamiltonian, when thought of as an operator on an n-fold tensor product space in the sense described above,  is a part of a hierarchy of conserved charges following from the solution of a Yang-Baxter equation.

The second problem  is related to the symmetries of the matrix model. To motivate the connection between the symmetries of the dilatation operator and the underlying Poisson algebra, we start by making the following observation.
In $H_{XXX}$, the two parts of the Hamiltonian commute with each other as operators on the Fock space, while in $H_{XXX-Matrix}$ that is not that case. In fact they do not seem to commute with each other in any of the obvious senses that one might think of. The strongest sense in which they might commute is in the sense of operators acting on the full Hilbert space of the matrix model, which amounts to the vanishing of their star commutator (\ref{star-mul}). But this is not true. 
\beq
\left[Tr (a^{\dagger i}a^{\dagger j}a_ia_j) \star Tr(a^{\dagger i}a^{\dagger j}a_ja_i)\right] \neq 0 
\eeq
One might imagine that they might commute in the sense of their large $N$ Poisson bracket(\ref{fullpoisson}) vanishing. But this turns out to be not true as well. i.e. 
\beq
\lim_{N \rightarrow \infty }\left\{ \frac{1}{\sqrt{N^{6}}} Tr (a^{\dagger i}a^{\dagger j}a_ia_j), \frac{1}{\sqrt{N^{6}}} Tr (a^{\dagger i}a^{\dagger j}a_ja_i)\right\}_1 \neq 0 
\eeq
A weaker condition might be that their Poisson bracket in the sense of $\hbar $ going to zero vanishes. By this we mean the Poisson bracket evaluated using the relation $\{ a_i , a^j \}_\hbar = \delta _i ^j 1\otimes 1$.  But even this, it turns out, is not true.
\beq
\left\{Tr (a^{\dagger i}a^{\dagger j}a_ia_j),Tr (a^{\dagger i}a^{\dagger j}a_ja_i)\right\}_\hbar \neq 0 \label{hbarclassical}
\eeq
In fact this lack of commutativity holds for the matrix models obtained from the higher charges of the spin chain as well. i.e the matrix models corresponding to all the higher charges do not commute with any part of the Hamiltonian or with each other in any of the three senses described above. However at the planar level, the matrix model is exactly equivalent to the spin chain, and, as one might expect, the two parts of the Hamiltonian commute with each other, if one uses the corresponding Poisson bracket (\ref{planar-lie}). This raises an interesting question. The planar Poisson bracket is a deformation (in the sense of $\hbar $) of the Poisson bracket given in (\ref{hbarclassical}). This quantization seems to produce an integrable system while its classical limit (\ref{hbarclassical}) is not! This clearly implies that the naive classical limit of the matrix model is not the correct  or the most useful one. For the purposes of integrability, the correct classical limit should then be the one obtained by letting both $\hbar $ and $\frac{1}{N}$ go to zero. This Poisson bracket, which we denote by double brackets has the following form.
 Let given two operators,
\beqs
\Theta ^I_J = \frac{1}{{\sqrt{N^{|I| +|J|-2}}}}Tr\left(a^{\dagger i_1}\cdots a^{\dagger i_{|I|}}a_{j_1}\cdots a_{j_{|J|}}\right),\nonumber \\
\Theta ^K_L =  \frac{1}{{\sqrt{N^{|K| +|L|-2}}}}Tr\left(a^{\dagger k_1}\cdots a^{\dagger k_{|K|}}a_{l_1}\cdots a_{l_{|L|}}\right).
\eeqs
 their Poisson bracket in the limit when both $\hbar $ and $\frac{1}{N}$ approach zero is,
\beqs
\left\{\{\Theta ^I_J,\Theta ^K_L\}\right\} = \delta ^{k_1}_{j_1}\Theta ^{Ik_2\cdots k_{|K|}}_{L j_2 \cdots j_{|J|}} + \delta^{k_{|K|}}_{j_{|J|}}\Theta ^{k_1 \cdots k_{|K|-1}I}_{ j_1 \cdots j_{|J|-1}L}\nonumber\\
-\delta ^{i_1}_{l_1}\Theta ^{Ki_2\cdots i_{|I|}}_{J l_2 \cdots l_{|L|}} 
-\delta^{i_{|I|}}_{l_{|L|}}\Theta ^{i_1 \cdots i_{|I|-1}K}_{ l_1 \cdots l_{|L|-1}J}.\label{classical}
\eeqs
One can evaluate, 
\beqs
\poisn{\Theta ^M_N}{\poisn{\Theta ^I_J}{\Theta ^K_L}} =\delta ^{k_1}_{j_1} \delta ^{i_1}_{n_1}\Theta ^{(M)(^\prime I)(^\prime K)}_{(L)(^\prime J)(^\prime N)} + 
\delta ^{k_1}_{j_1} \delta ^{k_{|K|}}_{n_{|N|}}\Theta ^{(I)(^\prime K ^\prime)(M)}_{(N^\prime)( L)(^\prime J)}\nonumber \\
-\delta ^{k_1}_{j_1} \delta ^{m_1}_{l_1}\Theta ^{(I)(^\prime K)(^\prime M)}_{(N)(^\prime L)(^\prime J)}
-\delta ^{k_1}_{j_1} \delta ^{m_{|M|}}_{j_{|J|}}\Theta ^{(M^\prime )(I)(^\prime K)}_{(L)( ^\prime J ^\prime )(N)}\nonumber\\
+\delta ^{k_{|K|}}_{j_{|J|}} \delta ^{k_{1}}_{n_{1}}\Theta ^{(M)(^\prime K ^\prime)(I)}_{(J^\prime)( L)(^\prime N)}+ 
\delta ^{k_{|K|}}_{j_{|J|}} \delta ^{i_{|I|}}_{n_{|N|}}\Theta ^{(K^\prime )(I ^\prime)(M)}_{(N^\prime)( J^\prime )(L)}\nonumber \\
-\delta ^{m_1}_{j_1} \delta ^{k_{|K|}}_{j_{|J|}}\Theta ^{(K^\prime )(I)(^\prime M)}_{(N)( ^\prime J ^\prime )(L)}-
\delta ^{k_{|K|}}_{j_{|J|}} \delta ^{m_{|M|}}_{l_{|L|}}\Theta ^{(M^\prime )(K ^\prime)(I)}_{(J^\prime)( L^\prime )(N)}\nonumber \\
-\delta ^{i_1}_{l_1} \delta ^{k_1}_{n_1}\Theta ^{(M)(^\prime K)(^\prime I)}_{(J)(^\prime L)(^\prime N)}
-\delta ^{i_1}_{l_1} \delta ^{i_{|I|}}_{n_{|N|}}\Theta ^{(K)(^\prime I ^\prime)(M)}_{(N^\prime)( J)(^\prime L)}\nonumber \\
+\delta ^{i_1}_{l_1} \delta ^{m_1}_{j_1}\Theta ^{(K)(^\prime I)(^\prime M)}_{(N)(^\prime J)(^\prime L)}
+\delta ^{i_1}_{l_1} \delta ^{m_{|M|}}_{l_{|L|}}\Theta ^{(M^\prime )(K)(^\prime I)}_{(J)( ^\prime L ^\prime )(N)}\nonumber \\
-\delta ^{i_{|I|}}_{l_{|L|}} \delta ^{i_{1}}_{n_{1}}\Theta ^{(M)(^\prime I ^\prime)(K)}_{(L^\prime)( J)(^\prime N)} - \delta ^{i_{|I|}}_{l_{|L|}} \delta ^{k_{|K|}}_{n_{|N|}}\Theta ^{(I^\prime )(K ^\prime)(M)}_{(N^\prime)( L^\prime )(J)}\nonumber \\
+\delta ^{m_1}_{l_1} \delta ^{i_{|I|}}_{l_{|L|}}\Theta ^{(I^\prime )(K)(^\prime M)}_{(N)( ^\prime L ^\prime )(J)}
+\delta ^{i_{|I|}}_{l_{|L|}} \delta ^{m_{|M|}}_{j_{|J|}}\Theta ^{(M^\prime )(I ^\prime)(K)}_{(L^\prime)( J^\prime )(N)},\nonumber \\
\eeqs
In the above equation, $I^{\prime }$ denotes the absence of an index, i.e. $I^ \prime = \{i_1 \cdots i_{(|I|-1)}\}$, $^\prime I = \{i_2 \cdots i_{(|I|)}\}$, $^\prime I^ \prime = \{i_2 \cdots i_{(|I|-1)}\}$etc.
Using the above relation, it is straightforward to show that,
\beq
\poisn{\Theta ^M_N}{\poisn{\Theta ^I_J}{\Theta ^K_L}} + \poisn{\Theta ^I_J}{\poisn{\Theta ^K_L}{\Theta ^M_N}} + \poisn{\Theta ^K_L}{\poisn{\Theta ^M_N}{\Theta ^I_J}} = 0
\eeq
Hence (\ref{classical}) defines a valid Poisson bracket. From the explicit $\hbar $ dependence given in (\ref{planar-lie}), it is easy to see that,
\beq
\poisn{\Theta ^I_J}{\Theta ^K_L} = \lim_{\hbar \rightarrow 0}\frac{1}{\hbar}\{\Theta ^I_J,\Theta ^K_L\}_2.
\eeq

A generic observable of the kind that we are interested in can be written as,
\beq
\Psi = \Psi_{i_1 \cdots i_{|I|}}^{j_{|J|} \cdots j_{|1|}}\Theta ^I_J.
\eeq
Note the reversal of order of the lower indices in the tensor $\Psi $.
For the Heisenberg model, the Hamiltonian is given by,
\beq
H_{XXX} \rightarrow \left(\Psi _2\right) ^{kl}_{ij} = \delta ^k_j\delta ^l_i - \delta ^k_i\delta ^l_j.
\eeq
The classical time evolution of observables on the Heisenberg model can be written is terms of the loop  variables as,
\beq
\dot{\Theta } ^K_L = \poisn{(\Theta _2) ^{kl}_{ij}}{ \Theta  ^K_L}
\eeq
The Hamiltonian has an obvious $su(2)$ invariance which is evident as,
\beq
\poisn{(\Theta _2) ^{kl}_{ij}}{ \Theta ^m_n \delta ^{ns}T_{sm}} = 0, \mbox{where $T$ is an $su(2)$ matrix.}
\eeq
As we shall see below, the symmetry is actually much larger. To be a generator of a symmetry transformation,  a classical observable must Poisson commute with the Hamiltonian with the Poisson structure given above. So identifying a set of observables that commute with each other and with the Hamiltonian with respect to this Poisson structure given above will be an indication of the presence of a bigger symmetry group. 

The first non-trivial case is that of a tensor of type $\chose{3}{3}$. The condition for a tensor of this type to commute with the Hamiltonian can be solved. A straightforward but laborious calculation gives,   
\beq
\left(\Psi _3 \right)^{p q r}_{i j k} = \delta ^{q}_{i}\delta ^{r}_{j}\delta ^{p}_{k}-\delta ^{r}_{i}\delta ^{p}_{j}\delta ^{q}_{k}
\eeq
as a solution. The corresponding matrix model operator 
\beq
(\Psi _3) = \frac{1}{N^2}Tr\left( a^{\dagger i}a^{\dagger j}a^{\dagger k}a_ja_ia_k - a^{\dagger k}a^{\dagger i}a^{\dagger j}a_ka_ja_i\right),
\eeq
maps to the spin chain operator $H_3 = \sum _l [P_{l,l+1},P_{l+1,l+2}]$, which is the next conserved charge in the Heisenberg Hierarchy. 


\begin{figure}
\centerline{\epsfxsize=4.truecm\epsfbox{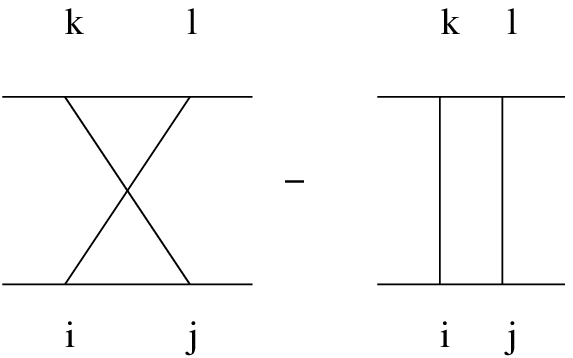}}
\caption{$H = P -I$}
\label{partition}
\end{figure}

\begin{figure}
\centerline{\epsfxsize=4.truecm\epsfbox{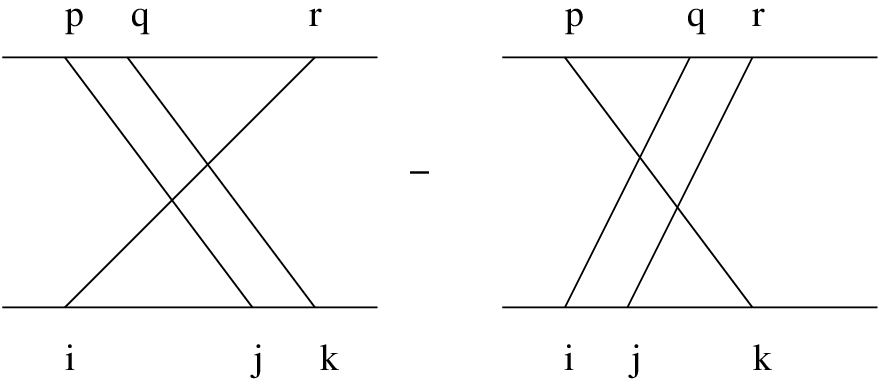}}
\caption{$[b_1,b_2]$}
\label{partition}
\end{figure}

\begin{figure}
\centerline{\epsfxsize=6.truecm\epsfbox{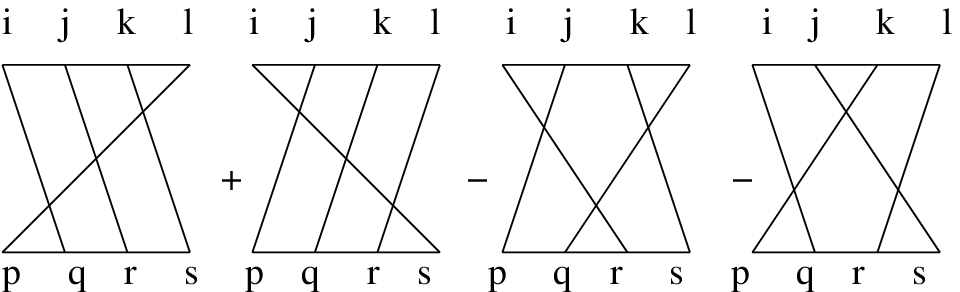}}
\caption{$[[b_1,b_2]b_3]$}
\label{partition}
\end{figure}
The process can be continued to hight orders as well. At the next order, one finds,
\beq
\left(\Psi _4\right)^{ijkl}_{pqrs} = \delta ^i _q\delta ^j _r\delta ^k _s \delta ^l _p + \delta ^i _s\delta ^j _p\delta ^k _q \delta ^l _r - \delta ^i _r\delta ^j _p\delta ^k _s \delta ^l _q - \delta ^i _q\delta ^j _s\delta ^k _p \delta ^l _r.
\eeq
upon quantization, this corresponds to,
\beq
H_4 = \sum_l \left[[P_{l,l+1},P_{l+1,l+2}]P_{l+2,l+3}\right]
\eeq
which is part of  Hamiltonian that one gets in the Heisenberg hierarchy which has interactions ranging up to four lattice sites. The charge we obtained is not  the complete Hamiltonian though, which is 
$\sum_l \left(\left[[P_{l,l+1},P_{l+1,l+2}]P_{l+2,l+3}\right] + P_{l,l+2}\right)$. The second piece is not present in the Hierarchy that we see in the classical limit. By continuing this process, one can see that the conserved charges of the classical limit of the Heisenberg chain constructed in this way correspond, upon quantization, to, 
\beq
H_n = \sum_l\left[\cdots[[P_{l,l+1},P_{l+1,l+2}],P_{l+2,l+3}] \cdots P_{l+(n-1)}\right]
\eeq 
A more detailed analysis of this classical limit and a proof of the classical charges being in involution will be presented elsewhere \cite{aa-inprep}.\\
{\bf Interpretation in terms of Braid Words:} This set of conserved charges can be indexed by a particular set of braid words.  One can diagrammatically denote the upper and lower sets of indices on the tensor $\Psi $, by two horizontal lines with the indices playing the role of lattice sites on these lines. The various $\delta $s can be represented by lines connecting the relevant lattice sites. The resultant diagram can be thought of as a braid diagram (figs 1,2 and 3), where the number of strands is equal to the number of upper and lower indices. We need not distinguish between over and under crossings at this level. The identity corresponds to the trivial braid on two strands. In fact, since with this Poisson bracket any tensor $\Psi ^I_I$ whose upper and lower indices coincide  generates an operator that commutes with everything else, any trivial braid on any number of strands will correspond to the identity. If we denote by $b_i$, the generator of the braid group that switches the lower end points of the $i$th  and $i+1$th strand keeping the top end points fixed, then the  permutation operator can be seen to correspond to the braid words $b_1$, $\Psi _3$ to $[b_1,b_2]$ and $ \Psi _4$ to $[[b_1,b_2],b_3]$. The $n$th charge corresponds to the word $[[\cdots[b_1,b_2],b_3],\cdots],b_{n-1}]$.\footnote{One might be tempted to think that this is an isomorphism between the algebra of braid words and the algebra of observables of matrix models with the Poisson bracket playing the role of a derivation on that algebra. This however is not the case.} 

Upon quantization this set does not go over to the complete set of charges of the spin chain. 
For a nice discussion on the nature of the conserved charges of the Heisenberg type of Hierarchies  see \cite{mathieu-charges-1}. It was shown there that the Hierarchy of Hamiltonians in the $XXX$ model has the structure of a Catalan tree. The $n$th Hamiltonian is given for instance by,
\beq
Q_n = F_{n,0} + \sum_{l=1}^{n/2 -1}\sum_{l=1}^{k}\alpha _{k,l}F_{n-2k,l}.
\eeq 
The notation needs requires a little explanation. $\alpha $'s are numerical coefficients, related to the generalized Catalan numbers $C_{l,m}$ as,
\beq
\alpha _{k,l} = C_{2k-l-1,l}, C_{n,m} = \chose{n-1}{p} - \chose{n-1}{p-2}.
\eeq
$F_{n,0}$, in the classical limit, corresponds to $H_n$, that we obtained by the analysis above. The other part of $Q_n$, can be understood as follows. Let us denote an increasing sequence of lattice sites $i_1,i_2 \cdots i_n$ by $C$. The Lattice sites are required to be in increasing order, but not necessarily nearest neighbors. One can think of the missing lattice points as 'holes'. Let,
\beq
f_n(C) = ((\cdots((\vec{\sigma }_{i_1}\times \vec{\sigma }_{i_2})\times\vec{\sigma }_{i_3})\cdots  )\times\vec{\sigma }_{i_{n-1}}).\vec{\sigma }_{i_{n}}.
\eeq
Denoting the set of all lattice increasing sequences beginning having $n$ points, with $k$ holes, by $C_{n,k}$,the expression for $F_{n,k}$ is \cite{mathieu-charges-1},
\beq
F_{n,k} = \sum_{C\in C_{n,k}}f_n(C).
\eeq
Hence, not all the charges of the quantum spin chain survive this classical limit. Or more precisely, for every Hamiltonian in the hierarchy of the quantum spin chain, there is a corresponding conserved quantity in this classical limit, which is just the $F_{n,0}$ part of $Q_n$. In other words, the part of the conserved charges that is built only out of neighboring spins is of $O(1)$, while the ones that involve holes are of $O(\hbar )$ and higher, and hence do not survive this classical limit. Needless to say, if one repeats the analysis with respect to the deformation of this Poisson structure with respect to $\hbar $, (\ref{planar-lie}), one will get the full spectrum of charges of the spin chain, which is a way of seeing that the quantization of this classical limit preserves integrability.

A few comments about this classical limit are in order here. This classical limit is different from the two well known integrable classical limits of the quantum spin chain. The first one is the naive classical  limit of the  spin chain, which has the  Hamiltonian $H = \sum S_i ^a S_{i+1}^a$, with the spins satisfying the canonical Poisson brackets (\ref{canpsn}). This, it turns out, is not a good classical limit, as the set of local conserved quantities corresponding to this Hamiltonian is not at all obvious. One can bypass this problem by going to the infinite volume limit. There the corresponding $1+1$ dimensional field theory Hamiltonian reads, 
\beq
H = \int _0^Ldx Tr(\partial_x S \partial_x S), S(x) = \sum_{i=1}^3S^i(x)\sigma _i, 
\eeq
along with the canonical Poisson brackets given by,
\beq
\{S^i(x),S^j(y)\} = \epsilon ^{ijk}S^k(x)\delta (x-y). 
\eeq

This system possesses Yangian symmetries, and the conserved quantities are non-local, as opposed to the ones we obtained above. The conserved quantities are given by the expansion of the transfer matrix,
\beq
T(x,\lambda ) = P e^{-\int _0 ^x \frac{i}{\lambda }S(y)}.
\eeq
A lattice version of these non-local charges can be constructed as well.
It is worth noting that this Yangian symmetry of the dilatation operator also has a counterpart in the Gauge theory, \cite{witten-yangian-1, witten-yangian-2}.
To get local conserved charges, one needs to modify the classical Hamiltonian, by replacing it with its logarithm.
\beq
H = \sum _j \ln(1 + S^a _jS^a _{j+1}),  
\eeq
with
\beq
\{S^i_a,S^j_b\} = \epsilon ^{ijk}S^k\delta _{ab} .\label{canpsn}
\eeq
This Hamiltonian does posses classical conserved quantities corresponding to the ones present in the quantum system. These problems related to the classical limits of Heisenberg chains are discussed in \cite{mathieu-charges-1, fadbook}.
The classical limit of the dilatation operator that we have here has some of the virtues of both of these limits. In the matrix model picture, the classical theory was obtained by the naive contraction of the Poisson structure of the corresponding quantum theory, which make it similar, in spirit, to the first case discussed above, but the presence of an infinite tower of local conserved charges make it similar, in content, to the second one.  
\subsection{Summary and Future Directions:} To reiterate the central point of the investigation, we see that the conserved charges of the quantum spin chain commute with each other and with the Hamiltonian with respect to the Lie bracket given in (\ref{planar-lie}). The corresponding classical charges display the same involutory properties with respect to the Poisson structure (\ref{classical}). Generalization of this approach to the full $SO(6)$ sector is reasonably straightforward using the work presented in \cite{minahan-spin, resh-1}. In the same sense, that the spin chain could be understood as a classical theory, the BMN limit too can be given a similar classical interpretation, and perhaps working out the plane wave limit of the dilatation operator in this approach can shed some light on symmetries present in that limit which are otherwise not obvious. 

As far as understanding the symmetries of the dilatation operator is concerned,  contraction of the planar Lie algebra and the resultant classical limit presented here plays a role similar to the one played by classical Yangian symmetries in the study of integrable field theories \cite{bern-1, clss-yang}.  It should be kept in mind though, that  the discussion here was completely within the context of local charges. We believe that this contraction  is potentially of some importance in understanding the classical limit of matrix theory and  we hope to report on this in the near future. 

{\bf Acknowledgments:} We would like to thank Ashok Das, Sumit Das, Horatiu Nastase and Radu Roiban for useful discussions. This work was supported in part by US Department of Energy grant number DE-FG02-91ER40685.

\section{Appendix 1:}
The canonical commutation relations obeyed by the basic observables of the matrix model are,
\beq
\left[ a^{\alpha i}_{j}, a^{\dagger \beta k}{_l} \right] = \hbar \delta^{\alpha \beta}\delta ^{i}_l \delta ^k _j.
\eeq
It can be cast in a more symmetric form using,
\beqs
A^\alpha = a^\alpha , \alpha = 1 \cdots 6 \nonumber \\ 
A^\alpha = a^{\dagger \alpha} , \alpha = 7 \cdots 12,
\eeqs
such that
\beqs
\left[ A^{\alpha i}_{j}, A^{ \beta k}{_l} \right] = \hbar \delta ^{i}_l \delta ^k _j \omega ^{\alpha \beta}\nonumber \\
\omega ^{\alpha \beta} = \canmat.
\eeqs

Given two normal ordered operator valued functions $f, g$ of the matrix valued creation and annihilation operators, their multiplication is isomorphic to the deformed product of the functions of the $A^\alpha  $'s thought of as ordinary matrices with the following start product.

\beq
f[A] \st g[A] = f[A] \multi g[A]\label{star-mul}
\eeq 
Traces of a string of the matrices, $Tr A^{i_1 }\cdots A^{i_n }$, can be regarded as functions of the matrices, or equivalently as functions indexed by the ordered set of indices $I = \{1_i \cdots i_n\}$.  Given two such functions, normalized as
\beqs
f[I] = \frac{1}{\sqrt{N^{|I| +2}}}Tr A^{i_1 }\cdots A^{i_n }, \nonumber\\
f[J] = \frac{1}{\sqrt{N^{|J| +2}}}Tr A^{j_1 }\cdots A^{j_n },
\eeqs 
their Poisson bracket in the Large $N$ limit is obtained by picking out the leading order terms in $\frac{1}{N}$ in the antisymmetric part of the star product given above \cite{Rajeevwilson}. 
\beqs
\{f[I],f[J]\}_{1} = \lim _{N\rightarrow \infty }N^2[f[I] \st f[J]]
= \sum_{r = 1}^{\infty }\sum_{\stackrel{\mu _1 < \mu _2 \cdots <\mu _r}{\nu _1 > \nu_2 \cdots >\nu _r}} \omega ^{i_{\mu _1}j_{\nu _1}}\cdots \omega ^{i_{\mu _r}j_{\nu _r}}\nonumber \\
f[I(\mu _1,\mu _2)J(\nu _2,\nu _1)] f[I(\mu _2,\mu _3)J(\nu _3,\nu _2)]\cdots f[I(\mu _r,\mu _r)J(\nu _1,\nu _r)]\label{fullpoisson} 
\eeqs
Further exposition on this Poisson bracket, can be found in \cite{rajeev-lee-review}.
This Poisson bracket allows one to realize the  large $N$ limit of any Hamiltonian matrix model as an interacting classical theory. Time evolution of observables being given by,
\beq
\frac{\partial }{\partial t}f[I] = \{H[M],f[I]\}_{1},
\eeq
where $H[M]$ is the Hamiltonian.  

It is quite evident from (\ref{fullpoisson}) that this Poisson bracket is highly non-linear. The non-linearities have to do with terms of rather high orders in $\hbar $ which survive the large $N$ limit. 
A linearization of this non-linear Poisson algebra may be obtained  by restricting the action of the observables  to low lying  excited states. This produces the  Lie algebra\cite{rajeev-lee-prl}(\ref{planar-lie}).  

{\bf Comments on Extracting the Divergent Part of Feynman Integrals and the Resulting 'Derivation':}\\ 
It is evident that the algebraic structure underlying the dynamics of the dilatation operator, thought of as a quantum mechanical system, is that of an associative algebra with a derivation on it. The associative algebra is which  the algebra of matrix valued creation and annihilation operators and the derivation being the commutator. In the large $N$ limit, this structure is replaced by that of a Poisson algebra. It is worth noting that this algebraic structure is not present in general in the full operator product expansion of the gauge theory, but manifests itself in the process of the extraction of the divergent parts of Feynman diagrams. In a sense, the process of extracting the divergent pieces of Feynman diagrams and subtracting away the divergences is very much like the process of differentiation. The relation of the dilatation generator to the divergent part of operator product expansions was pointed out in \cite{beisert-et-al-conformal}.
For the sake of completeness we shall briefly repeat their argument here. One first notices that the tree level correlation function can be written as,
\beq
\left<\Upsilon ^{I_1\cdots I_n}(x)\Upsilon ^{J_1\cdots J_m}(y)\right>_{Tree} = e^{\Delta _{xy}Tr\frac{\partial}{\partial \Phi _i(x)}\frac{\partial}{\partial \Phi _i(y)}}\Upsilon ^{I_1\cdots I_n}(x)\Upsilon ^{J_1\cdots J_m}(y)|_{\Phi = 0}
\eeq
At higher loops, this result is modified by the replacing the tree level propagator 
\beq
\Delta _{xy} = \frac{\Gamma (1-\epsilon )}{|\frac{1}{2}(x-y)^2|^{1-\epsilon }},
\eeq
by the corresponding renormalized propagator. After renormalizing the composite operators, the one loop, the  expression for the two point function is,
\beq
\left<\Upsilon (x)\Upsilon (y)\right>_{One-Loop} =  e^{\Delta _{xy}Tr\frac{\partial}{\partial \Phi _i(x)}\frac{\partial}{\partial \Phi _i(y)}}\left(1 + g^2 V_2(x) -g^2V_2(x_0)\right)\Upsilon (x)\Upsilon (y)|_{\Phi = 0}.
\eeq
For compactness, we have dropped the indices specifying the traces in the above equation. $V_2$ above, is the one loop effective vertex, which encodes the sum of the various Feynman diagrams contributing to the corelation functions at this order. For the scalar sector, the relevant one loop diagrams correspond to gluon exchange, scalar vertex insertion and self energy corrections to the scalar propagator. $x_0$ is the scale. The effective vertices are divergent, but the difference  $V_2(x) -g^2V_2(x_o)$ is finite. This finite difference is the dilatation operator,
\beq
\lim_{\epsilon \rightarrow 0}\left(V_2(x) -g^2V_2(x_o)\right) = \ln(x_0^2/x^2)\Gamma.
\eeq 
So the manifestly finite form of the one loop correlator is,
\beq
\left<\Upsilon (x)\Upsilon (y)\right>_{One-Loop}= e^{\Delta _{xy}Tr\frac{\partial}{\partial \Phi _i(x)}\frac{\partial}{\partial \Phi _i(y)}}e^{\ln(x_0^2/x^2)\Gamma_{x}}\Upsilon (x)\Upsilon (y)|_{\Phi = 0}.
\eeq
The subscript on $\Gamma $ implies that it only acts on the fields at $x$. From the discussion above, we know that this action involves a Poisson algebra. Hence the process of 'differentiation ' manifests itself as a derivation on the algebra underlying the dynamics of the dilatation generator. 

\bibliography{abhishekbib}

\end{document}